\documentclass{appolb}
\usepackage{epsfig}
\begin{document}
\title{Cooling of superfluid neutron stars with muons}
\author{M. Bejger
\address{N.\ Copernicus Astronomical Center, \\
	  Bartycka 18, 00-716 Warsaw, Poland \\
          {\tt bejger@camk.edu.pl}},
\and
D. G. Yakovlev
\address{ Ioffe Physical Technical Institute, \\	
           Politekhnicheskaya 26, 194021 St.~Petersburg, Russia \\
           {\tt yak@astro.ioffe.rssi.ru}},
\and
O. Y. Gnedin
\address{ Space Telescope Science Institute, \\
	 3700 San Martin Drive, Baltimore, MD 21218, USA \\
         {\tt ognedin@stsci.edu}}
}
\maketitle
\begin{abstract}
We extend our modeling of cooling of superfluid neutron stars
by including the production of muons in the core, in addition to neutrons,
protons, and electrons. The results are confronted with observations
of middle-aged isolated NSs. Muons have little effect on the hydrostatic
structure of NSs, on the slow cooling of low-mass NSs (RX J0822--43
and PSR 1055--52 in our model) and on the rapid cooling
of massive NSs. They affect, however, 
the moderately fast cooling of medium-mass NSs
(1E 1207--52, RX J0002+62, PSR 0656+14, Vela, and Geminga) and shift
appreciably the mass range of these NSs to lower masses, which is important
for correct interpretation of the observations. 
Moreover, the effects of muons can accurately be
reproduced by a simple renormalization of NS models
with no muons in the NS cores.
\end{abstract}

\PACS{97.60.jd, 67.90, 65}
  
\section{Introduction}
 
Rapid progress
in detecting thermal emission from
isolated neutron stars (NSs)
with a new generation of orbital and ground-based
observatories (e.g., Refs.\ \cite{pavlovetal02,pzs02})
stimulates active theoretical studies
of cooling isolated NSs. It is well known that
cooling history of NSs depends on 
physical properties of matter
of supranuclear density in NS cores.
These properties (composition of matter,
equation of state -- EOS, 
critical temperatures of various superfluids,
etc.) are largely unknown: they cannot be reproduced
in laboratory or unambiguously calculated
(because of the lack of exact manybody theory
for describing the systems
of particles interacting via strong forces).
However, they may be constrained by
comparing cooling calculations with observations
of isolated NSs (e.g., Ref.\ \cite{page98}).

We will focus on the theoretical interpretation
of observations proposed recently in Refs.\ \cite{khy01,ykg01},
\cite{kyg02} (hereafter KYG),
and \cite{ykhg02,ygkp02}.
The authors restricted themselves to
model EOSs in the
NS cores in which the presence of muons was
neglected. In the present paper we show that the effect
of muons on the cooling may be important.

For the observational basis, we take the same isolated
middle-aged NSs as KYG, but exclude RX J1856-3754.    
The age of this intriguing radio-quiet NS has recently been
revised \cite{wl02}; the present value $t=5 \times 10^5$ yrs
is lower than the former one, $9 \times 10^5$ yrs,
and the source has become less restrictive
for cooling theories (cf.\ KYG and
Ref.\ \cite{ykhg02}). In addition, there are indications
that the emission from the entire NS surface is obscured
by the emission from a hot spot on the surface of the NS; if so
the surface temperature is poorly determined from
the present observations (e.g., \cite{pzs02} and references therein).

The effective surface
temperatures, $T_{\rm s}^\infty$,
redshifted for a distant 
observer, and ages $t$ of seven isolated NSs
are taken from Table 3 of KYG
and displayed in Figs.\ 2 and 3 below.
The three youngest objects,
RX J0822--43 \cite{ztp99}, 1E 1207--52 \cite{zpt98},
and RX J0002+62 \cite{zp99}, are radio-quiet
NSs in supernova remnants.

The other objects,
Vela (PSR 0833--45)\ \cite{pavlovetal01},
PSR 0656+14 \cite{pmc96},
Geminga (PSR 0633+1748)\ \cite{hw97},
and PSR 1055--52 \cite{ogelman95},   
are observed as radio pulsars.       
The adopted values of $T_{\rm s}^\infty$ are inferred
from the observed spectra using various models of stellar
emission described in KYG.
Recently, the values of   
$T_{\rm s}^\infty$ for some of the sources have been
revisited in Refs.\ \cite{pavlovetal02,pzs02},
\cite{sanwaletal02}--
\cite{sandroetal02}. 
Since the new data are basically in line with those
used in KYG, we do not introduce the corresponding 
changes; they are insignificant for our analysis.  

As shown in KYG, the observations can be
explained using the models of NSs with the
cores composed only of neutrons, protons, and
electrons, and assuming the presence of nucleon
superfluidity with the density dependent critical
temperatures $T_{\rm c}(\rho)$.
Following KYG we consider
superfluidities of three types produced by: (1) singlet-state
pairing of protons in the NS core ($T_{\rm c}= T_{\rm cp}$); 
(2) singlet-state pairing of free neutrons in the
inner crust ($T_{\rm c}=T_{\rm cns}$); and
(3) triplet-state pairing of neutrons in  
the core ($T_{\rm c}=T_{\rm cnt}$).
Owing to a large scatter of microscopic theoretical models
of $T_{\rm c}(\rho)$ 
(e.g. Ref.\ \cite{ls01}),
we treat $T_{\rm c}(\rho)$ as
free parameters.

KYG considered cooling of NSs with rather
strong pairing of protons and
weak pairing of neutrons in the core,
and with a strong pairing of neutrons in the crust.
They found that cooling middle-aged
NSs can be divided into three types.
\begin{description}
\item[$\bullet$]   
Type I NSs are low-mass NSs which show
slow cooling with
(modified or direct) Urca processes of neutrino emission
strongly suppressed by proton superfluidity.
The cooling curves, $T_{\rm s}^\infty(t)$,  
are insensitive to NS mass, EOS
in the core, and proton superfluidity
(i.e., to $T_{\rm cp}$)
as long as the latter is sufficiently strong.
KYG interpreted RX J0822--43, and PSR 1055--52
as low-mass NSs.

\item[$\bullet$]
Type II NSs are medium-mass NSs which show
moderately fast cooling regulated by direct Urca process
partly reduced by proton superfluidity
in the NS central kernels. The cooling curves
are sensitive to NS mass, EOS, and especially
the $T_{\rm cp}(\rho)$ profiles in the NS kernel.
If the EOS and $T_{\rm cp}(\rho)$ are fixed,
the effective surface temperature   
decreases smoothly with increasing $M$,
and one can measure the mass (`weigh'' medium-mass NSs)
using the observed limits on $T_{\rm s}^\infty(t)$.
KYG treated 1E 1207--52, RX J0002+62, Vela, 
PSR 0656+14, and Geminga as medium-mass NSs.

\item[$\bullet$]
Type III NSs are massive NSs which show rapid cooling
via direct Urca process in the NS kernels, almost unaffected by
proton superfluidity.
The surface temperatures of these NSs
are low (a few times $10^5$ K for $t \simeq 10^3$ yrs),
being not too sensitive to the NS structure.
No NS of such type has been observed so far.
\end{description}

\section{Cooling models}

We use the same cooling code
as in KYG and modify the physics input
in the NS core to include the effects of muons.

First, we have included muons in the EOS.
We use a 
stiff EOS proposed in Ref.\ \cite{pal88}, the model I of the symmetry
energy of nucleon matter
with the compression modulus of saturated
nuclear matter $K=240$ MeV.  The {\it same model of nucleon-nucleon
interaction} was adopted by KYG (EOS A in their notations)
who, however, artificially suppressed
the presence of muons. Now we include the muons
and obtain EOS A$\mu$. We will compare
the results obtained with EOSs A and A$\mu$.

For EOS A$\mu$, the muons appear at $\rho \ge \rho_\mu =
2.55 \times 10^{14}$ g cm$^{-3}$ (when the electron
chemical potential exceeds the muon rest energy).  
Their fraction is lower than 10\% everywhere
in the NS core. Their appearance
slightly softens the EOS, slightly increases the
fraction of protons and decreases the fraction of electrons.
These changes weakly affect the NS internal structure,
reducing the maximum NS mass by 1.4\%.
The masses $M$, central densities $\rho_{\rm c}$, and radii $R$
of two NS configurations for EOSs A and A$\mu$
are given in Table 1. The first configuration corresponds
to a maximum-mass NS. The second
corresponds to the onset of powerful direct Urca neutrino
emission \cite{lpph91}
in the NS kernel ($\rho_{\rm c}=\rho_{\rm D}$,
$M=M_{\rm D}$).

\begin{table}[t]
\centering
\caption{NS models employing EOSs A and A$\mu$ (without and with muons)}

\begin{tabular}{|l|l|l|l|l|l|}
\hline
Model  & Main parameters                       &     EOS A	&  EOS
A$\mu$\\
\hline \hline
Maximum& $M_{\rm max}/{\rm M}_\odot$		&  1.977		&
1.950    \\
mass   & $\rho_{\rm cmax}/10^{14}$ g cm$^{-3}$	&  25.75		&
26.55    \\
model  & $R$, km                                &  10.754	 &  10.602
\\
\hline \hline
Direct Urca& $M_{\rm D}/{\rm M}_\odot$        &   1.358		&  1.249
\\
threshold& $\rho_{\rm D}/10^{14}$ g cm$^{-3}$	&   7.851		&
7.423    \\
model              & $R$, km                   &   12.98	 &  12.952
\\
\hline
\end{tabular}
\end{table}  

Another modification of the cooling code consists of incorporating
the heat capacity of muons. It
is straightforward (e.g., Ref.\ \cite{yls99})
and has no noticeable effect on the NS cooling.

Finally, we have modified the emissivities of
neutrino reactions in the NS cores (as described
in Ref.\ \cite{ykgh01}).
The main effect of muons
is to {\it lower the threshold density $\rho_{\rm D}$ of
onset of direct Urca process} (Table 1). The lowering is associated with
the changes of the fractions of neutrons, protons, and electrons
in muonic matter which relaxes the onset condition.
At $\rho < \rho_{\rm D}$,
modified Urca process is basically the leading one.
At $\rho > \rho_\mu$,
two new branches of this process
appear, the neutron and proton branches,
where the muons participate instead of the electrons.
However, these new branches have
little effect on the cooling because their emissivity
is comparable to
the emissivity of the ordinary modified Urca
process.
The ordinary direct Urca process
operates at $\rho > \rho_{\rm D}$.
At still higher density $\rho > \rho_{\rm D\mu}$
($\rho_{\rm D\mu} = 9.257 \times 10^{14}$ g cm$^{-3}$,
for EOS A$\mu$) another direct Urca
process is open, 
where the muons participate instead
of the electrons. We have included it into the code but,
again, it has little effect on the cooling because its emissivity
is comparable to the emissivity of the ordinary 
direct Urca process. 

\begin{figure}
\begin{center}
\leavevmode   
\epsfxsize=84mm
\epsffile[20 143 575 695]{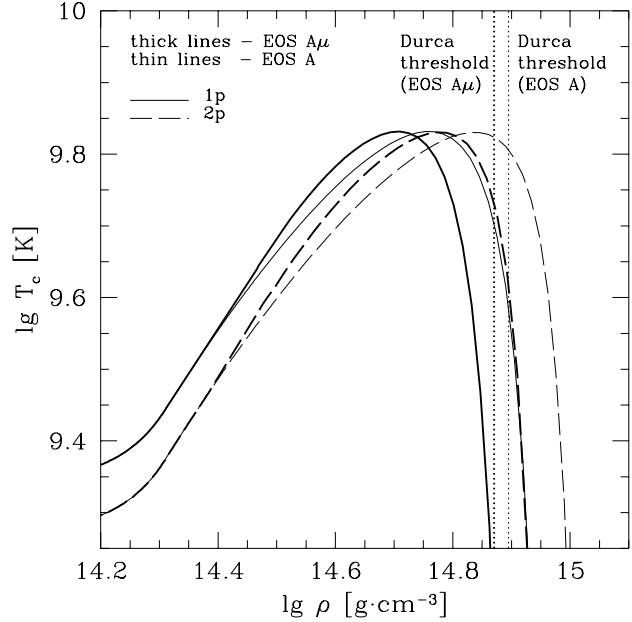}
\end{center}
\caption{   
Density dependence of the critical temperature
of proton superfluidity for models 1p and 2p, and
for 
EOS A$\mu$ (with muons, thick lines) and
EOS A (without muons, thin lines).
Vertical dotted lines indicate
the direct Urca (Durca) threshold.
}
\label{fig1}
\end{figure}

Now we describe our models of superfluidity in NSs.
As in KYG, we assume that
the triplet-state pairing of neutrons in the NS cores
is weak (maximum $T_{\rm cnt} < 10^8$ K) and
can be ignored.  Otherwise,
a strong neutrino emission due to Cooper
pairing of neutrons would accelerate the cooling
and hamper the interpretation of observations of older NSs.
The singlet-state pairing of neutrons in the crust
controls the neutrino luminosity of low-mass
NSs and has no direct relation to the presence
of muons in the NS core. While exploring the effects of muons we
ignore this pairing (see also Sect.\ 3).

\begin{figure}
\centering
\epsfxsize=86mm
\epsffile[20 143 570 700]{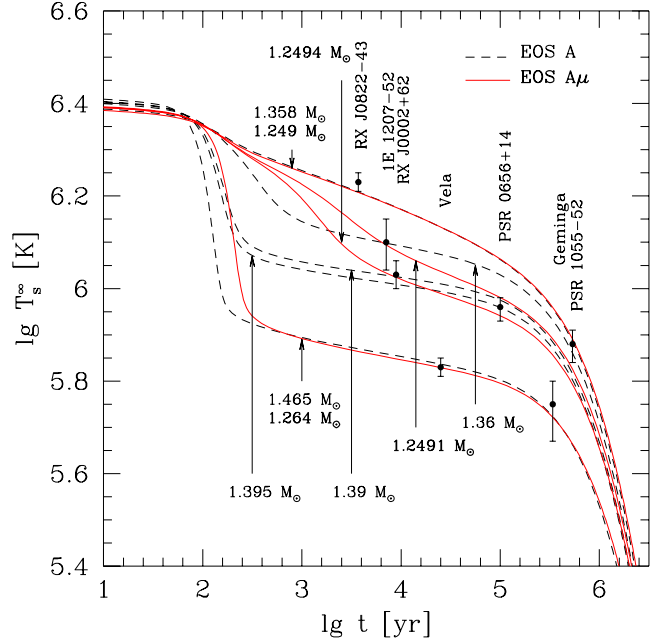}
\caption{
Observational limits on surface temperatures of seven
NSs compared with cooling curves
for NSs of several masses, adopting
1p proton superfluidity and either 
EOS A$\mu$ (with muons, solid lines)
or ESO A (without muons, dashed lines).
If two masses are given, an
upper value corresponds to EOS A and
a lower (boldface) value to EOS A$\mu$.
}
\label{fig2}
\end{figure}

Thus, we focus on the proton
superfluidity in the core.  
We parameterize the dependence of $T_{\rm cp}$
on the proton Fermi wavenumber $k=(3 \pi^2 n_{\rm p})^{1/3}$
($n_{\rm p}$ being the proton number density) 
by the same expression as Eq.\ (1) in KYG
and consider two models of superfluidity,
1p and 2p, employed by KYG.
The corresponding $T_{\rm cp}(\rho)$
profiles are plotted in Fig.\ \ref{fig1}.
The maximum of $T_{\rm cp}(\rho)$ is about the same
($\sim 7 \times 10^9$ K) for all models,
but superfluidity 2p extends to a higher densities than 1p.
Since we use the {\it same model of nucleon interaction}   
for EOSs A and A$\mu$, we have the {\it same
dependence of} $T_{\rm cp}$ {\it on} $k$ (but different
dependences of $T_{\rm cp}$ on $\rho$ because the
proton fractions are different for EOSs A and A$\mu$).
The curves in Fig.\ 1 are
typical of the microscopic calculations
(e.g., Ref.\ \cite{ls01}).
As discussed above,
the muons shift both the direct Urca threshold and the $T_{\rm cp}(\rho)$
profile to lower values of $\rho$.
The latter shift, due to the increase in the proton fraction, is bigger.
For instance, $T_{\rm cp}(\rho)$ for model 1p 
of proton superfluidity vanishes below
the direct Urca threshold in the muonic matter, but
persists above the direct Urca threshold
in the matter without muons.

\section{Results and discussion}

\begin{figure}
\centering
\epsfxsize=86mm
\epsffile[20 143 570 700]{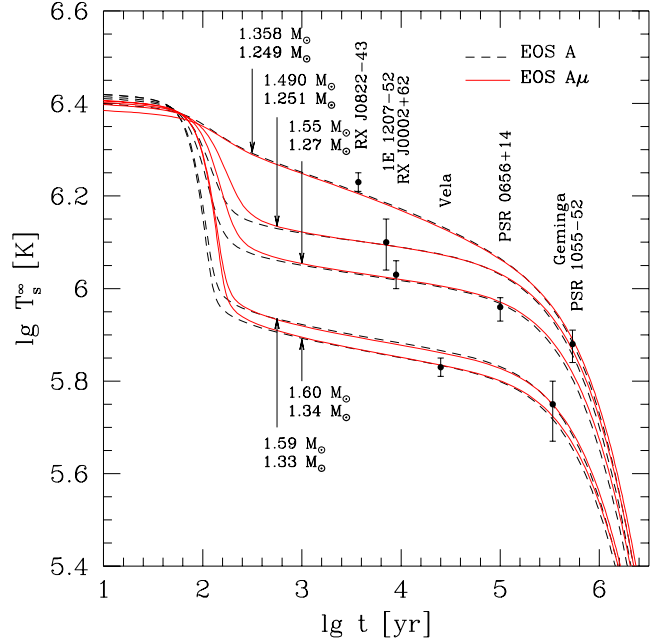}
\caption{
Same as Fig.\ 2, but for proton superfluidity model 2p.
}
\label{fig3}
\end{figure}

The results of the cooling calculations are illustrated
in Figs.\ \ref{fig2} and \ref{fig3}, which also show
the observational data.  

According to our models, the cooling curves
of low-mass NSs with and without muons are almost
indistinguishable. The neutron superfluidity 
in the crust can affect the cooling of these stars
in exactly the same way as discussed
in Ref.\ \cite{ykg01} and KYG
for the NSs without muons.   
The calculations also show that muons
have almost no effect on rapid cooling of
massive NSs.

However the effect of muons on moderately fast
cooling of medium-mass NSs is substantial.
Figure 2 corresponds to model 1p of proton superfluidity.
Without muons, the $T_{\rm cp}(\rho)$ profile extends
above the direct Urca threshold (Fig.\ 1).
Accordingly, we have a representative class of
medium-mass NSs (dashed curves).  By varying the mass
$M$ from 1.36 ${\rm M}_\odot$ to 1.465 ${\rm M}_\odot$
we can explain the observations of the five sources (1E 1207--52,
RX J0002+62, Vela, PSR 0656+14, and Geminga) by the cooling
of medium-mass NSs, and thus `weigh''
these NSs (Ref.\ \cite{khy01}, KYG).  
In the presence of muons, the proton critical temperature
$T_{\rm cp}(\rho)$ drops to zero below the
direct Urca threshold, $\rho_{\rm D}$ (Fig.\ 1), i.e.,
the proton superfluidity disappears
in the centers of NSs with masses somewhat lower
than the threshold mass $M_{\rm D}$. The transition
from slow to fast cooling occurs in a very narrow  
mass range, from 1.249 M$_\odot$ to 1.264 M$_\odot$,
so we have no representative class of
medium-mass NSs.  The proposed interpretation of the
five sources as medium-mass NSs containing muons becomes unlikely
(just as in the absence of superfluidity, see KYG).

Figure 3 is the same as Fig.\ 2
but for model 2p of proton superfluidity.
This superfluidity extends above
$\rho_{\rm D}$ with and
without muons (Fig.\ 1). Therefore, we have
representative classes of medium-mass NSs  
and we can weigh our five medium-mass sources
in both cases. Without muons, as in KYG,
we obtain the masses from 1.49 M$_\odot$ to
1.60 M$_\odot$. With muons,
we obtain significantly lower (and narrower)
mass range for the same sources, from
1.27 M$_\odot$ to 1.34 M$_\odot$.

The inferred mass range changes in the presence of
muons because they
lower the threshold density $\rho_{\rm D}$
of direct Urca process and shift the decreasing slope
of the $T_{\rm cp}(\rho)$ profile to an even lower density
(Fig.\ 1).  Accordingly, direct Urca
process becomes stronger at lower $\rho$ and this
accelerates the cooling of lower-mass NSs.
Therefore, the mass range of medium-mass  
NSs shifts to lower $M$.

\section{Conclusions}

We have
simulated the cooling of superfluid NSs
with and without muons.
The inclusion of muons 
does not violate the main results of
Refs.\ \cite{khy01}--\cite{ygkp02}  
on the existence of three types of cooling
superfluid NSs.
As in KYG, we can interpret
RX J0822--43, and PSR 1055--52,
as low-mass NSs, and 
1E 1207--52, RX J0002+62, Vela, PSR 0656+14, and Geminga,
as medium-mass NSs. This interpretation 
requires strong singlet-state proton pairing
and weak triplet-state neutron pairing in the NS cores.

Muons do not change the hydrostatic structure of NSs.
The muon heat capacity and direct/modified
Urca processes involving muons
have almost no effect
on NS cooling.  The cooling curves
of low-mass and high-mass NSs are insensitive
to the presence of muons.

However, muons \emph{do affect}
the cooling of medium-mass
NSs and the associated `weighing'' of observed NSs
proposed in Ref.\ \cite{khy01} and KYG.
For the same nucleon-nucleon interaction model
and the superfluid properties of NS cores,
we obtain noticeably lower masses of NSs with
muons than without muons (Fig.\ 3).
Therefore, one should take muons into account
for a correct interpretation of
observations.

It is remarkable that muons \emph{affect the cooling by their
mere existence, as passive particles}. We could switch
off all neutrino reactions and heat capacity associated
with muons (i.e., retain the `old'' physics input
in the NS core)
but employ the fractions of neutrons, protons and electrons
in the EOS modified by the presence of muons.
The cooling curves obtained in this numerical experiment
would almost coincide with the exact ones.
This means that  \emph{the effects of muons can be accurately
incorporated by renormalizing the NS models
containing no muons at all}. In other words,
any model of a cooling NS with muons is equivalent
to a certain model of a cooling NS without muons. 
Clearly, the muons will have similar effects
in the presence of hyperons in supranuclear matter.
This circumstance may be helpful in constructing the
codes to simulate cooling of NSs containing hyperons.

The existence of three types of cooling NSs
is effectively regulated by two factors.   
The first is the onset of the powerful     
direct Urca process of neutrino emission   
at densities $\rho > \rho_{\rm D}$;
the second is the strong proton superfluidity
extending at $\rho > \rho_{\rm D}$.  The latter suppresses
the direct and modified Urca processes at or just above   
the critical density
$\rho_{\rm D}$ but opens them at higher $\rho$.
In this way the superfluidity smears out the sharp
transition between the slow and rapid cooling regimes
and creates a representative
class of medium-mass NSs, favorable for the interpretation
of observations.  These conditions
may be also realized 
in NS cores containing hyperons,
pion or kaon condensates, or quark matter.
We will analyze 
different models of NS cores in future
publications.

\par{
We are grateful to G.G.~Pavlov for encouragement,
to D.\ Page for criticism, and to
P.\ Haensel and A.D.\ Kaminker for careful reading of the text.
The work was partially supported by RFBR (grant No.\ 00-07-90183)
and KBN (grant No.\ 5P03D.020.20).
}


\begin{thebibliography}{}

\bibitem[1]{pavlovetal02}
Pavlov, G.~G., Sanwal, D., Garmire, G.~P., \& Zavlin, V.~E.,
{\it Neutron Stars in Supernova Remnants},
eds.\ P.~O.\ Slane, B.~M.\ Gaensler,
{\it ASP Conf.\ Ser.} (2002) p.\ 247

\bibitem[2]{pzs02}
Pavlov, G.~G., Zavlin, V.~E., \& Sanwal, D.,
{\it Proc.\ of 270 Heraeus Seminar on Neutron Stars,   
Pulsars and Supernova Remnants},
eds.\ W.\ Becker, H.\ Lesh, \& Tr\"umper, MPE, Garching
(2002) submitted [astro-ph/0206024]

\bibitem[3]{page98}
Page, D.,
{\it The Many Faces of Neutron Stars},
eds.\ R.\ Buccheri, J.\ van Paradijs, \& M.~A.\ Alpar,
Kluwer, Dordrecht (1998), p. 539

\bibitem[4]{khy01}
Kaminker, A.~D., Haensel, P., \& Yakovlev, D.~G.,
{\it A\&A} {\bf 373}, L17, (2001)

\bibitem[5]{ykg01}
Yakovlev, D.~G., Kaminker, A.~D., \& Gnedin, O.~Y.
{\it A\&A} {\bf 379}, L5, (2001)

\bibitem[6]{kyg02}
Kaminker, A.~D., Yakovlev, D.~G., \& Gnedin, O.~Y.,
{\it A\&A } {\bf 383}, 1076 (2002)

\bibitem[7]{ykhg02}
Yakovlev, D.~G., Kaminker, A.~D., Haensel, P., \& Gnedin, O.~Y.,
{\it A\&A } {\bf 389}, L24 (2002)

\bibitem[8]{ygkp02}
Yakovlev, D.~G., Gnedin, O.~Y., Kaminker, A.~D., \& Potekhin, A.~Y.
{\it Proc.\ of 270 Heraeus Seminar on Neutron Stars,
Pulsars and Supernova Remnants},
eds.\ W.\ Becker, H.\ Lesh, \& Tr\"umper, MPE, Garching
(2002) submitted [astro-ph/0204226]  

\bibitem[9]{wl02}
Walter, F.~M., \& Lattimer,~J.,
{\it ApJ (Letters)} (2002) submitted [astro-ph/0204199]

\bibitem[10]{ztp99}
Zavlin, V.~E., Tr\"{u}mper, J., \& Pavlov, G.~G.,
{\it ApJ} {\bf 525}, 959, (1999)

\bibitem[11]{zpt98}
Zavlin, V.~E., Pavlov, G.~G., \& Tr\"{u}mper, J.,
{\it A\&A} {\bf 331}, 821 (1998)

\bibitem[12]{zp99}
Zavlin, V.~E., \& Pavlov, G.~G.,
private communication (1999)

\bibitem[13]{ponsetal01}
Pons, J., Walter, F., Lattimer, J., Prakash, M.,
Neuh\"{a}user, R., \& An, P., {\it ApJ} {\bf 564}, 981 (2002)

\bibitem[14]{pavlovetal01}
Pavlov, G.~G., Zavlin, V.~E., Sanwal, D., Burwitz, V., \& Garmire, G.~P.,
{\it ApJ}  {\bf 552}, L129 (2001)

\bibitem[15]{pmc96} Possenti, A., Mereghetti, S., \& Colpi, M.,
{\it A\&A} {\bf 313}, 565 (1996)

\bibitem[16]{hw97}
Halpern, J.~P., \& Wang, F.~Y.-H.,
{\it ApJ} {\bf 477}, 905 (1997)   

\bibitem[17]{ogelman95} \"{O}gelman, H.,
{\it Lives of Neutron Stars},
eds.\ M.~A.\ Alpar, \"U.\ Kizilo\u{g}lu, J.\ van Paradjis,
{\it NATO ASI Ser.\ } (Kluwer, Dordrecht) {\bf p.\ 101 } (1995)

\bibitem[18]{sanwaletal02}
Sanwal, D., Pavlov, G.~G., Kargaltsev, O.~Y., Garmire, G.~P.,
Zavlin, V.~E., Burwitz, V., Manchester, R.~N., Dodson, R.,   
{\it Neutron Stars in Supernova Remnants},
eds.\ P.~O.\ Slane, B.~M.\ Gaensler,      
{\it ASP Conf.\ Ser.} (2002) [accepted, astro-ph/0112164]

\bibitem[19]{burwitzetal01}
Burwitz, V., Zavlin, V.~E., Neuh\"auser, R., Predehl, R.,
Tr\"umper, J., \& Brinkman, A.~C., 
{\it A\&A Lett.} {\bf 327}, L35 (2001)

\bibitem[20]{gbr01}
G\"ansicke, B.~T., Braje, T.~M., \& Romani, R.~W.,
{\it A\&A} {\bf 386} 1001 (2002) 

\bibitem[21]{kva01}
Kaplan, D.~L., van Kerkwijk, M.~H., \& Anderson, J.,
{\it ApJ} {\bf 571} 447 (2002) 

\bibitem[22]{pzst02}
Pavlov,~G.~G., Zavlin,~V.~E., Sanwal,~D., \& Tr\"umper,~J.,
{\it ApL (Letters)} {\bf 569} L95 (2002)

\bibitem[23]{ms02}
Marshall,~H.~L., \& Schulz,~N.~S.,
{\it ApJ} {\bf 574} 377 (2002)

\bibitem[24]{sandroetal02}
Mereghetti,~S., De Luca,~A., Caraveo,~P.~A., Becker,~W.,
Mignani,~R., \& Bignami,~G.~F.,
{\it ApJ (Letters)} submitted (2002) [astro-ph/0207296]

\bibitem[25]{ls01}
Lombardo, U., \& Schulze, H.-J.,
{\it Physics of Neutron Star Interiors},
eds.\ D.\ Blaschke, N.\ Glendenning, A.\ Sedrakian
(Springer, Berlin) {\bf p.\ 30} 

\bibitem[26]{pal88}
Prakash, M., Ainsworth, T.~L., \& Lattimer, J.~M.,
{\it Phys.\ Rev.\ Lett.} {\bf 61}, 2518 (1988)

\bibitem[27]{lpph91}
Lattimer, J.~M., Pethick, C.~J.,  Prakash, M., \& Haensel, P.,
{\it Phys.\ Rev.\ Lett.}  {\bf 66}, 2701 (1991)

\bibitem[28]{yls99}
Yakovlev, D.~G., Levenfish, K.~P., \& Shibanov, Yu.~A.,
{\it Physics--Uspekhi} {\bf 42}, 737 (1999) [astro-ph/9906456]

\bibitem[29]{ykgh01}
Yakovlev, D.~G., Kaminker, A.~D., Gnedin, O.~Y., \& Haensel, P.,
{\it Phys.\ Rep.} {\bf 354}, 1 (2001)

\end{thebibliography}
\end{document}